\documentclass[aps,prb,showpacs,twocolumn,showpacs]{revtex4-1}

\usepackage{graphicx}
\usepackage{amsmath}
\usepackage{amssymb}
\usepackage{wasysym}
\usepackage{textcomp}
\usepackage{color}

\definecolor{lila}{rgb}{0.5,0,1}
\newcommand{\bnen}{\begin{equation}}
\newcommand{\eden}{\end{equation}}
\newcommand{\bean}{\begin{eqnarray}}
\newcommand{\eean}{\end{eqnarray}}

\newcommand{\bna}{\begin{array}}
\newcommand{\eda}{\end{array}}

 \definecolor{darkgreen}{rgb}{0,0.5,0} 
 \usepackage{color}
 \usepackage[normalem]{ulem}
 \usepackage{cancel}
\begin{document}

\title{Entanglement and magnetism in high-spin graphene nanodisks}
\author{I. Hagym\'asi\thanks{hagymasi.imre@wigner.mta.hu} }
\affiliation{Department of Physics and Arnold Sommerfeld Center for Theoretical Physics, Ludwig-Maximilians-Universität München, Theresienstrasse 37, 80333 München, Germany}
\affiliation{Strongly Correlated Systems "Lend\"ulet" Research Group, Institute for Solid State
Physics and Optics, MTA Wigner Research Centre for Physics, Budapest H-1525 P.O. Box 49, Hungary}
\author{\"O. Legeza}
\affiliation{Strongly Correlated Systems "Lend\"ulet" Research Group, Institute for Solid State
Physics and Optics, MTA Wigner Research Centre for Physics, Budapest H-1525 P.O. Box 49, Hungary}

\begin{abstract}
We investigate the ground-state properties of triangular graphene nanoflakes with 
zigzag edge 
configurations. The description of zero-dimensional nanostructures 
requires accurate many-body techniques since the widely used density-functional theory with local 
density approximation or Hartree-Fock methods cannot handle the strong quantum fluctuations.
Applying the unbiased density-matrix renormalization group algorithm we 
calculate the magnetization and entanglement patterns with high accuracy
for different interaction strengths and compare them to the mean-field results. With the help of 
quantum information analysis and subsystem density matrices we reveal that the edges are strongly 
entangled with each other. We also address the effect of electron and hole doping and demonstrate 
that the 
magnetic properties of triangular nanoflakes can be controlled by electric field, which 
reveals features of flat-band ferromagnetism.
This may open up 
new avenues in graphene based spintronics.

\end{abstract}
\maketitle
\section{Introduction}
Up to lately a common stereotype has been that magnetism is primarily attributed to $d$ 
electron 
systems. Nowadays more and more supporting evidence emerges that $sp$ electron systems can host 
also magnetic moments, moreover, the $sp$ magnetism is predicted 
to be stable even at room temperature \cite{Makarova2006vii}. 
After the observation of magnetic moments in defective samples\cite{Esquinazi:graphite,Nair2012},
enormous attention has been paid to monolayer graphene to investigate if it can exhibit 
long-range magnetic order. The emergence of long-range magnetic order  in graphene  is expected 
along the zigzag edges of samples due to their peculiar nature.
The existence of such magnetic order has 
been controversial until the appearance of modern nanofabrication methods due to the poor edge 
quality in experiments. The recent development of top-bottom techniques allows now us to tailor 
graphene samples with 
atomic precision\cite{Tapaszto:litography}, and the bottom-up synthesis is capable of 
creating even smaller nanostructures\cite{bottom-up1,PhysRevLett.100.056807,Ruffieux2016}. 
The most studied nanostructures with zigzag edges are the graphene nanoribbons, hexagonal
and triangular nanoflakes. 
It has been demonstrated that nanoribbons can be 
accessed by scanning-tunneling 
litography\cite{Tapaszto:litography} and bottom-up synthesis as well\cite{Kimouche2015}. The 
theoretically 
predicted\cite{louie:prl2006,Louie:GW,PhysRevLett.101.096402,Rossier:HF,Yazyev:prl2008,
MacDonald:prb2009,Feldner:prl2011,Hagymasi:dmrg} magnetism of nanoribbons with zigzag edges 
has been corroborated by several indirect experimental 
evidence\cite{Wang2016,Magda2014,Ruffieux2016} since direct observation of magnetic moments would 
require a macroscopic quantity of nanoribbons which is beyond the scope of the present techniques.
The fast development of bottom-up techniques triggered the investigation of other geometries, like 
the hexagonal and triangular nanoflakes. Very recently, a nanoflake close to a triangular
shape has been created with chemical vapour deposition.\cite{Hagymasi:prb2017} The
hexagonal structure consists of equal number of atoms from both 
sublattice, while the triangular one is, however, an uncompensated lattice, which has
attracted 
significant attention recently.\cite{Rossier:prl2007,PhysRevLett.102.157201,Potasz:prb2010} This 
interest is due to the topological frustration\cite{PhysRevLett.102.157201} caused by the sublattice 
imbalance which leads to a ground-state degeneracy proportional to the system 
size.\cite{Potasz:prb2010} The triangular structure is also interesting from another point of view, 
namely, it has a ground state with nonzero spin, whose magnitude is proportional to the sublattice 
imbalance according to 
Lieb's theorem.\cite{Lieb:1989} Thus, unlike in compensated lattices (graphene nanoribbons, 
hexagonal nanoflakes), where magnetism occurs beyond certain sizes of the zigzag segments, 
triangular systems are 
expected to be magnetic for all sizes. The possibility that they can host net magnetic moments has 
inspired active research in this field. The magnetic properties of these nanodisks have been 
explored by various techniques, including Hartree-Fock\cite{Rossier:prl2007},
density-functional theory (DFT),\cite{Wang:nanolett}  and 
configuration-interaction methods.\cite{PhysRevLett.103.246805,PhysRevB.85.075431} In
low-dimensional systems correlation effects become important whose proper treatment
requires accurate many-body techniques. Besides a few relevant studies for
nanoribbons\cite{Schmidt:QMC,Feldner:prl2011,Hagymasi:dmrg} and hexagonal
structures\cite{Feldner:prb2010,Valli:prb2016}, such an analysis for triangular structures
is still lacking.
\par Our goal in this paper is to fill this gap by performing large-scale numerical simulations with
the unbiased density-matrix renormalization group algorithm (DMRG).\cite{White:DMRG1}
With the help of the
true many-body ground state we can analyze its entanglement structure, furthermore, its
total spin can be directly assessed,\cite{PhysRevA.83.012508} which cannot be done in DFT or 
mean-field
calculations. We check the reliability of mean-field theory for nanodisks by comparing the 
magnetization to the DMRG results. Furthermore, we compare the ground-state properties away from 
half-filling with those obtained from configuration interaction methods.
\par The paper is organized as follows. In Sec.~II.~ we provide the details of our DMRG
calculations and describe the main steps of the mean-field approach. In Sec.~III.~we
briefly recall the properties of the noninteracting system and in Sec.~III.~ A our DMRG results are 
presented using the elements
of quantum information theory, while in Sec.~III.~B the magnetic properties are discussed
in the half-filled case and compared to the mean-field results. In Sec.~III.~C we 
address the role of hole and electron doping and show that the system displays the features 
of a flat-band ferromagnetism also away from half band filling.\cite{mielke1993} Finally, 
in Sec.~IV.~ we conclude our results.

\section{Methods}
We consider the widely used $\pi$-band model of graphene to describe the triangular
quantum dot including a local Hubbard-interaction term,
\begin{equation}
\label{eq:Hamiltonian}
\mathcal{H}=-t\sum_{\langle
ij\rangle}\hat{c}^{\dagger}_i\hat{c}^{\phantom\dagger}_j+U\sum_i\hat{n } _ { i\uparrow}
 \hat{n}_{i\downarrow},
\end{equation}
where $t=2.7$ eV is the nearest-neighbor hopping amplitude, and
$U$ is the strength of the 
local Coulomb interaction. In spite of its simplicity, it has been shown that the Hubbard model 
with properly chosen parameters and fillings can quantitatively reproduce experimental 
results in graphene systems.\cite{Magda2014,Hagymasi:prb2017} However, less 
attention has been 
paid to the model's properties on a triangular nanoflake.
\par To gain some physical insight into the properties of a triangular nanoflake, we 
apply the
mean-field approach first. By neglecting 
the fluctuation terms in the Hamiltonian
(\ref{eq:Hamiltonian}), we obtain an effective single-particle Hamiltonian
\begin{equation}
 \label{eq:Hamiltonian_mf}
 \mathcal{H}_{\rm MF}=-t\sum_{\langle ij
\rangle\sigma}\hat{c}^{\dagger}_{i\sigma}\hat{c}^{\phantom\dagger}_{j\sigma}+U\sum_{
i\sigma}\langle \hat{n}_{i\bar{\sigma}} \rangle \hat{n}_{i\sigma},
\end{equation}
where the unknown electron densities, $\langle \hat{n}_{i\bar{\sigma}}\rangle$ are determined
by using the standard self-consistent procedure.
 \par To account for the quantum fluctuations and many-body effects, we use the real-space DMRG 
algorithm \cite{White:DMRG1,White:DMRG2,schollwock2005,manmana2005,hallberg2006,legeza:review}.
We map our short-ranged 2D Hamiltonian to a one-dimensional chain topology with long-range 
couplings   (see Sec.~III, 
Fig.~\ref{fig:mutual_inf_U0_zz}). 
The area law limits the available system sizes since the DMRG cost scales 
exponentially with the entanglement entropy which is proportional to the size of the triangle.
In order to decrease the truncation errors to the order of $10^{-4}$, we
kept up to 20000 block states. With such a large bond dimension, we were able to determine 
the 
many-body ground state and correlation functions accurately.
\section{Results}
\begin{figure}[!t]
\includegraphics[width=.8\columnwidth]{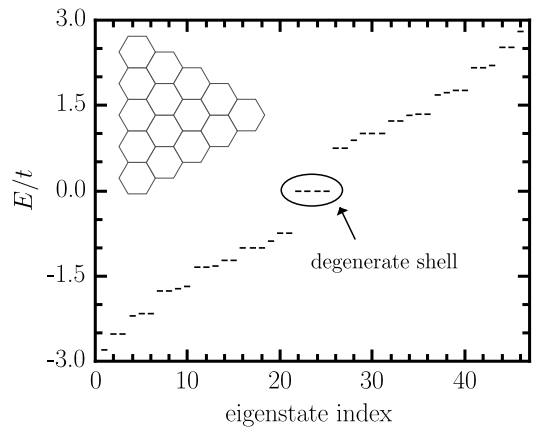}
\caption{ The energy spectrum of the noninteracting system. The inset shows the 
investigated system, the circled levels form the degenerate shell at the Fermi level.}
\label{fig:spectrum}
\end{figure}

We start by briefly recalling the properties of the noninteracting case. The system we 
consider is seen in the inset of Fig.~\ref{fig:spectrum}.
Just like in zigzag nanoribbons zero-energy states appear at the Fermi energy, but here they form a 
degenerate shell due to the absence of translational symmetry.\cite{Ezawa:prb2007}
Their degeneracy is $N_{\rm edge}-1$ according to the general theorem in 
triangular nanodisks,\cite{Potasz:prb2010} where $N_{\rm edge}$ is the number of edge atoms along a 
side of the triangle.  These degenerate states are sensitive to electron-electron interaction and 
play a crucial role -- like in the fractional quantum Hall effect\cite{PhysRevLett.48.1559}  -- 
and are responsible for the flat-band magnetism\cite{mielke1993} occurring in the system.
\subsection{Quantum information analysis, correlation functions}
\begin{figure}[!t]
\includegraphics[width=0.9\columnwidth]{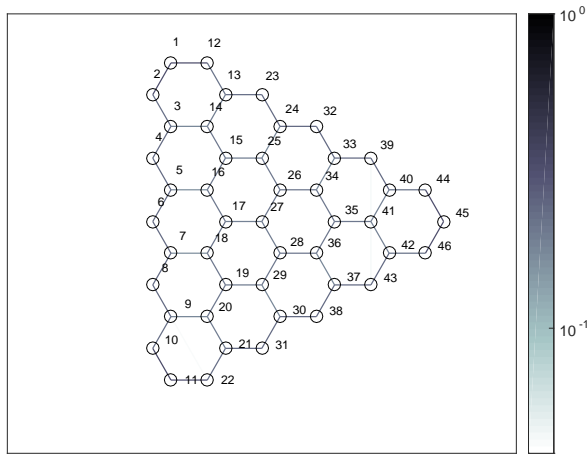}
\caption{ Entanglement patterns in a zigzag nanodisk for $U=0$. The
magnitude of the mutual information components is encoded using the grayscale in the
sidebar. The numbers indicate the positions 
of sites along the one-dimensional DMRG topology.}
\label{fig:mutual_inf_U0_zz}
\end{figure}

\begin{figure}[!t]
\includegraphics[width=0.9\columnwidth]{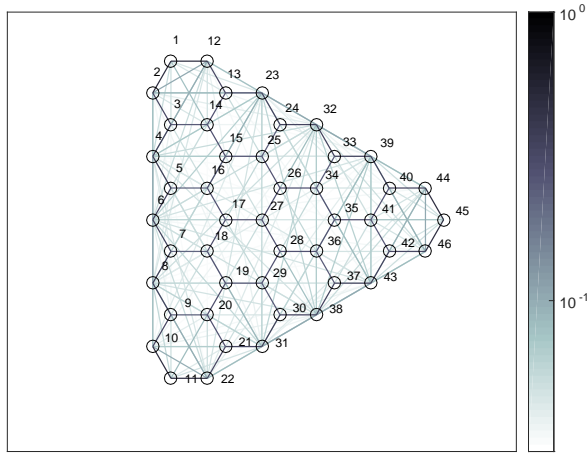}
\caption{ Similar to Fig.~\ref{fig:mutual_inf_U0_zz} but for $U/t=2$.}
\label{fig:mutual_inf_U2_zz}
\end{figure}

\begin{figure}[!ht]
\includegraphics[width=\columnwidth]{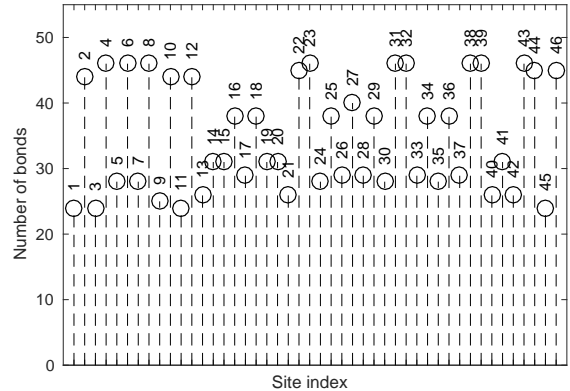}
\caption{The degree of lattice sites in Fig.~\ref{fig:mutual_inf_U2_zz} taking into
account mutual information bonds larger than $10^{-2}$. }
\label{fig:NOB}
\end{figure}

 First, we examine the ground-state properties of the half-filled system by
calculating various correlation 
functions. We study the correlations between two arbitrary sites within our system, which can be 
characterized by 
the mutual information, \cite{wolf2008,furukawa2009,legeza:entanglement}
\begin{gather}
   I_{ij}=s_i+s_j-s_{ij}.
\end{gather}
It measures all types of correlations (both of classical and quantum origin) between sites $i$ 
and $j$. This quantity is often referred as the strength of entanglement between the two sites 
embedded in the whole system. 
Here $s_i$ and $s_{ij}$ are the one- and two-site von Neumann 
entropies, \cite{legeza2003b,vidallatorre03,calabrese04,legeza2006,rissler2006,luigi2008}
respectively, that can be calculated from the corresponding one- and two-site reduced density 
matrices,
\begin{align}
 s_i&=-{\rm Tr} \rho_i\ln\rho_i,\\
 s_{ij}&=-{\rm Tr} \rho_{ij}\ln\rho_{ij},
\end{align}
 where $\rho_i$ ($\rho_{ij}$) is the reduced density matrix of site $i$ (sites $i$ and $j$),
which is obtained from the density matrix of the total system by tracing out the configurations of
all other sites. 
In what follows, we explore the entanglement structure of the ground state by calculating the mutual 
information for different 
values of the Hubbard interaction to investigate the role of the electron-electron
interaction.
We begin with the noninteracting case, the entanglement structure within the nanodisk is
shown in Fig.~\ref{fig:mutual_inf_U0_zz} for $U=0$.
It is immediately seen that only short-ranged correlation develop between the sites,
which is typical for noninteracting systems. This drastically changes as the interaction
is switched on, as is seen in Fig.~\ref{fig:mutual_inf_U2_zz} for $U/t=2$.
We can observe
that the sites along each zigzag edge becomes correlated, moreover, long-ranged
entanglement appears between any two edges. This is somewhat similar to what happens in
zigzag nanoribbons, but there are some important differences which will be explained in
the following. The fact that the edge sites are the most entangled can be visualized by
considering the correlation values about a given threshold ($10^{-2}$) and counting the
degree of each site. As it can be easily seen from Fig.~\ref{fig:NOB} the edge sites that belong to 
the zigzag-edge sublattice
possess the largest degree. 
\par The mutual information provides us an overall picture about which sites are
strongly correlated with each other. To reveal its origin, it is instructive to
investigate the eigensystem of the corresponding two-site density matrices. As a first
step, we consider two neighboring zigzag sites at the edge, 
(4 and 6 in Fig.~\ref{fig:mutual_inf_U2_zz}), and solve the eigenvalue problem of the 
corresponding two-site reduced density matrix, $\rho_{4,6}$. In its eigenvalue spectrum,
the most 
significant eigenvalue ($\omega=0.13$) is threefold degenerate, and the corresponding
eigenvectors 
are: \begin{equation}
\label{eq:wavefunction:V0p3_onsite}
\begin{split}
 \phi^{(1)}_{4,6}=& \ |\uparrow\rangle_{4}|\uparrow\rangle_{6},\\
 \phi^{(2)}_{4,6}=& \
\frac{1}{\sqrt{2}}(|\uparrow\rangle_{4}|\downarrow\rangle_{6}+|
\downarrow\rangle_{4}|\uparrow\rangle_{6}\rangle),\\
 \phi^{(3)}_{4,6}=& \ |\downarrow\rangle_{4}|\downarrow\rangle_{6}.
\end{split}
\end{equation}
Thus, the electrons in sites 4 and 6 form a triplet. Qualitatively, we have the same
result between every pair of sites along a zigzag edge, therefore ferromagnetic
correlation emerges at the edges, which is the usual behavior that one expects. It is
more interesting to perform this analysis for a pair of sites that are on two adjacent edges,
for example 4 and 32. A similar analysis of $\rho_{4,32}$ yields that we have the same
eigensystem for the most significant eigenvalue ($\omega=0.12$) as in the previous case.
Based on the strong mutual information between every pair of edges, we can conclude that
strong ferromagnetic coupling arises between the edges that has not been reported before.
Note that a similar scenario occurs in nanoribbons, namely, the correlations between sites 
along a zigzag edge are ferromagnetic, too, but there the coupling between the two edges is 
antiferromagnetic. The reason is that in the nanoflake the edge sites consist of the same 
sublattice sites, while in nanoribbons
the sites of two zigzag edges belong to the two different sublattices.
The nearest-neighbor correlation remains antiferromagnetic in the nanodisk too, which can be seen 
immediately from the analysis of $\rho_{4,5}$, where the eigenvector belonging to the dominating 
eigenvalue $\omega=0.42$ is a singlet:
\begin{equation}
\label{eq:wavefunction:V0p8_nn}
\begin{split}
 &\phi_{4,5}=\\
  &0.59(|\uparrow\rangle_{4}|\downarrow\rangle_{5}
 -|
\downarrow\rangle_{4}|\uparrow\rangle_{5})\\
& + 0.39(|\uparrow\downarrow\rangle_{4}|0\rangle_{5}
+|0\rangle_{4}
|\uparrow\downarrow\rangle_{5}).
\end{split}
\end{equation}
The result is qualitatively the same for any nearest-neighbor pair.
\begin{figure*}[!ht]
\includegraphics[width=1.3\columnwidth]{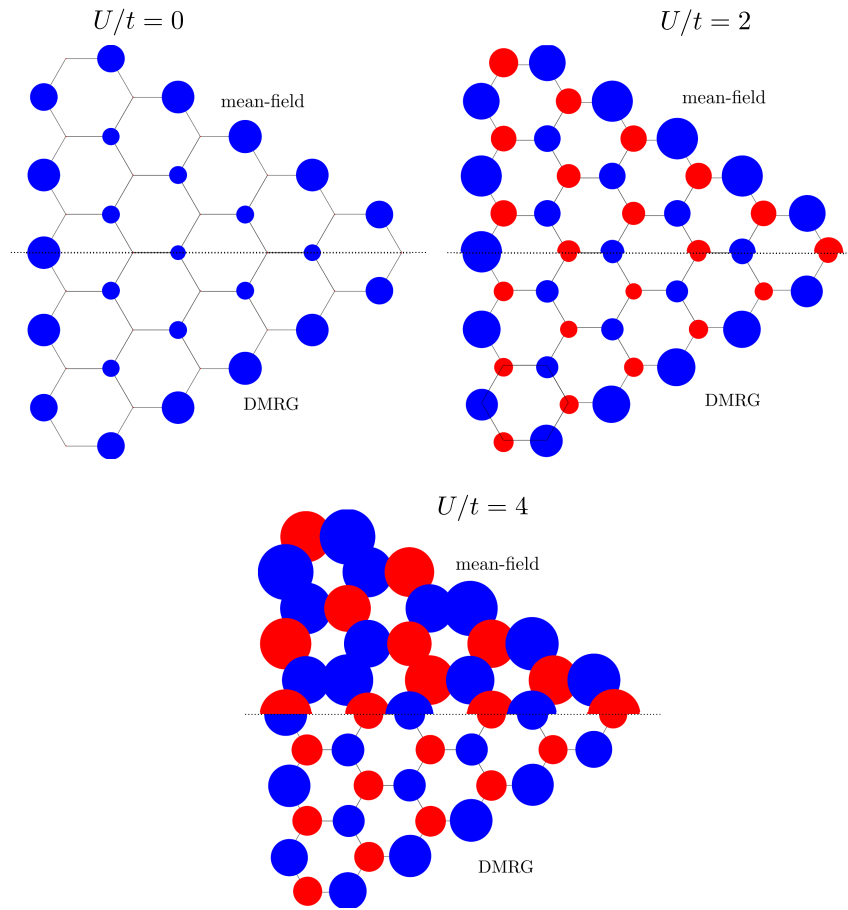}
\caption{ The local magnetic moments ($S_i^z$) of the ground for 
various values of $U$.  The magnitude of up (blue) and down (red) moments 
are proportional to the area of the circles.   }
\label{fig:triangle_magnetic}
\end{figure*}
\subsection{Magnetic properties at half filling}

In the previous subsection the behavior of correlation functions were studied, but -- as
mentioned in the Introduction~-- the magnetism of triangular nanodisks is especially 
interesting since even 
as small system as ours can exhibit net magnetic moments. We calculate the total spin of the ground 
state using the spin-spin correlation functions:
\begin{equation}
 \langle S^2\rangle=\left\langle\frac{1}{2}\sum_{ij}\left(S^+_iS^-_j+S^-_iS^+_j 
\right)+\sum_{ij}S^z_iS^z_j\right\rangle.
\end{equation}
In the half-filled case we find that the 
ground state has always $S=2$ spin for $U>0$ as it is dictated by the sublattice imbalance 
($N_A-N_B=4$) and 
Lieb's theorem ($N_A-N_B=2S$). However, the spin distribution is far from trivial and depends 
strongly on the Hubbard $U$. Since the mean-field theory is widely used for the description of 
graphene nanostructures, we check its predictions for the magnetization. 

The results from both 
methods are shown in Fig.~\ref{fig:triangle_magnetic} for various values of the Hubbard interaction.
In the noninteracting case, the ground state is not uniqe, therefore we choose the $S^z_{\rm 
tot}=2$ sector to be able to compare the magnetization with the DMRG result. At $U=0$ both methods 
give naturally the same result, namely, only the sites of 
sublattice $A$ (zigzag edge sites, shown by the blue circles) are polarized. Sublattice $B$ can be 
identified where $\langle S^z_i\rangle=0$, which was found to be of the order of $10^{-3}-10^{-4}$ 
in our 
DMRG calculations. Switching on $U$ results in the enhancement of nearest-neighbor 
antiferromagnetic correlations, and as a result, finite magnetization appears on the sites of 
sublattice $B$. It is remarkable that the mean-field theory gives quite close results to those of 
DMRG regarding the magnetization on sublattice $A$ (the difference is around 15\%), while it 
significantly overestimates the polarization on sublattice $B$, by a factor of 2. The edges 
carry 
 most of the net magnetic moments. For strong Coulomb interaction, $U/t=4$, the 
mean-field 
theory predicts large polarization inside the nanodisk like along the edges and their values are 
grossly overestimated compared to the DMRG results. The failure of the mean-field theory can be 
traced back to the fact that the infinite honeycomb lattice becomes antiferromagnetic above 
$U_c/t\sim2.2$ in the mean-field approach, and our calculation reflects this tendency. The DMRG 
results also show enhanced magnetization inside the nanoflake, however, their values are still less 
than the moments appearing at the edges. Larger values of $U$ for graphene are unphysical, since 
at $U_c/t\sim3.9$ a Mott transition occurs in  the two-dimensional honeycomb 
lattice.\cite{Mott1:prx,Mott2:prb,Mott3:prx}

\subsection{Magnetic properties for electron and hole doping}

\begin{figure}[!ht]
\includegraphics[width=.8\columnwidth]{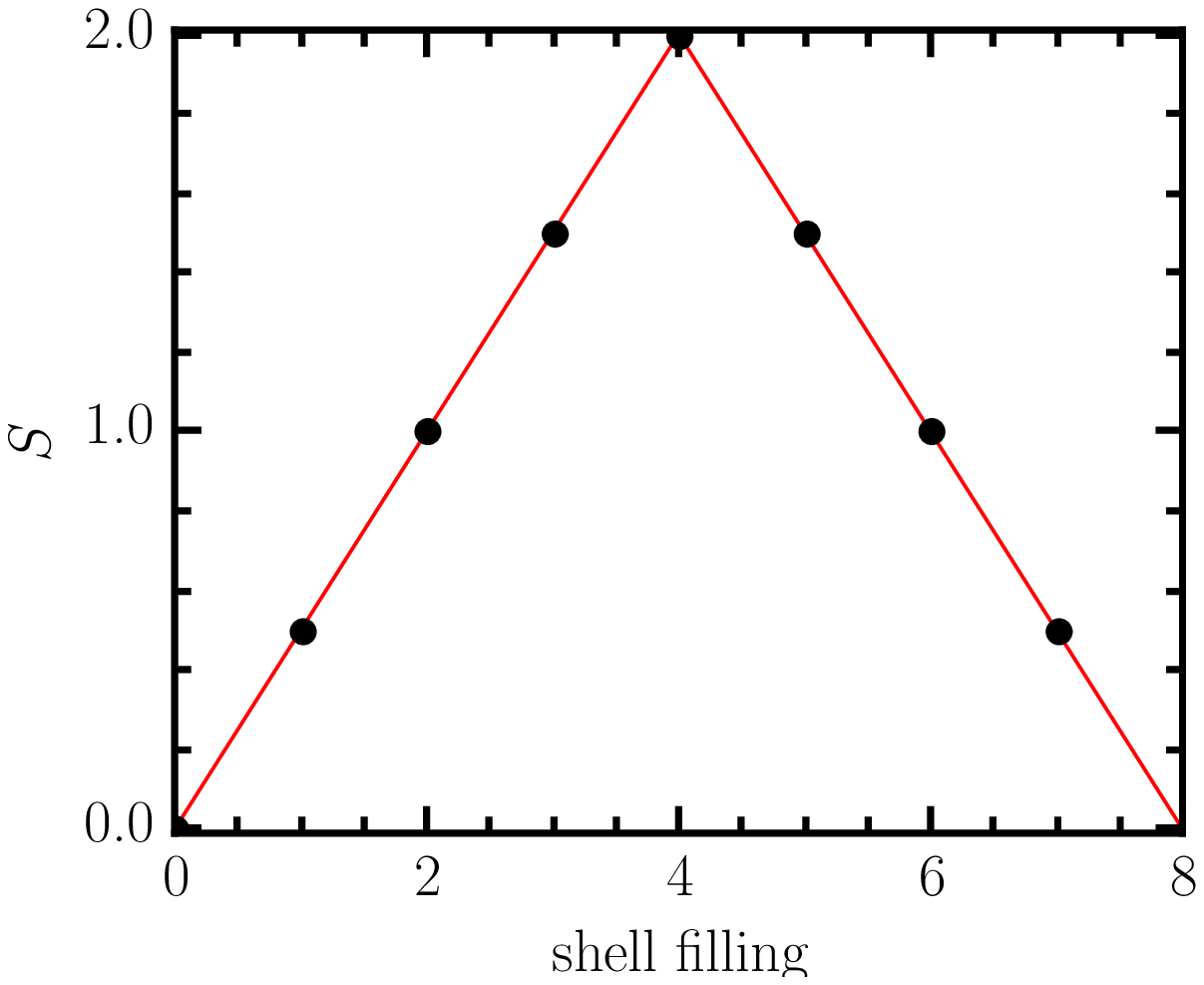}
\caption{The ground-state spin for $U/t=2$ as a function of filling of the fourfold degenerate 
shell in Fig.\ref{fig:spectrum}. The line is guide to the eye. The numerical error of $S$ was 
within the size of the symbols.}
\label{fig:filling}
\end{figure}

Lastly, we address the effect of hole and electron doping in 
the triangular nanoflake. While Lieb's theorem clearly determines the spin of the ground state in 
the half-filled case, it cannot be used away from half-filling. Previous calculations based on the 
configuration interaction method pointed out that doping can crucially affect the spin of the 
ground state.\cite{PhysRevB.85.075431,PhysRevLett.103.246805} Our findings are 
summarized in Fig.~\ref{fig:filling}, where the total spin is plotted against the filling of the 
fourfold degenerate shell in Fig.\ref{fig:spectrum}.

We find that the maximum spin belongs to the half-filled case, and it gradually 
decreases as we 
move away from half filling and it vanishes as soon as the shell is completely filled or becomes 
empty.
What is common with the previous results\cite{PhysRevB.85.075431,PhysRevLett.103.246805} is that 
(i) 
the maximum spin corresponds to the half-filled case, (ii) the nanoflake exhibits magnetization as 
long as the degenerate shell is partially filled and it disappears if it is fully occupied or no 
electrons occupy the shell. The configuration interaction method predicts a monotonous decrease of 
the spin for hole doping, just 
like in our case, however, for electron doping a non-monotonous behavior is
predicted. Namely, when 
5 electrons occupy the shell (this means adding a single electron to the half-filled system), a 
strong depolarization occurs and the ground state has a spin of 
$S=1/2$. In contrast, our calculation yields $S=3/2$ for the same filling. 
Our results are supported by Mielke's and Tasaki's results for 'flat-band ferromagnetism'.\cite{mielke1993}  In 
degenerate systems,
e.g., in partially filled atomic shells, the ground state has maximal spin (Hund's rule) because  a 
ferromagnetic aliagement
of the electrons' spins minimizes the Coulomb repulsion at no cost of the kinetic energy.  This 
immediately leads to the pattern
seen in Fig.~\ref{fig:filling}.
 Our results suggest that the triangular nanoflakes may be used as building blocks 
for spintronic devices,  the magnetism is 
being stable against doping and no complete depolarization occurs which would limit their 
usability.
Since doping can be controlled by an external electric field, it makes the 
triangular nanoflake an ideal candidate for spintronic applications.
\section{Conclusions}
In this paper we examined ground-state properties of triangular graphene nanoflakes 
by 
performing large-scale numerical calculations with the unbiased DMRG method. After a short revisit 
of the noninteracting case, we use the elements of quantum information theory to reveal that 
strongly entangled edge states emerge in this system. Its source can be attributed to the 
long-range ferromagnetic correlations between the edges that have not been pointed out so far. We 
also examined the magnetic properties of the triangular nanodisk and compared the magnetization 
values to those of mean-field results for various values of the Hubbard interaction. It turned out 
that for $U/t\sim 2$ the mean-field theory gives fairly good estimates for the edge magnetization, 
while it overestimates the magnetization in the bulk. Close to the Mott transition, the mean-field 
approach results in completely wrong magnetization values which is due to the enhanced quantum 
fluctuations. In each case we obtained a quintet ground state in agreement with  Lieb's theorem. 

Finally, we considered the effect of electron and hole doping in the system. By calculating the 
spin correlation functions, the ground-state spin could be unambiguously determined for all 
possible fillings of the degenerate shell. We found that the ground-state spin decreases gradually 
as the system is doped away from half-filling and the ground state becomes singlet for a completely 
empty or full shell, in agreement with the prediction of flat band ferromagnetism.  
In particular, the high-spin 
ground state is preserved in the sense that no complete depolarization occurs for small doping.

\acknowledgements{ We acknowledge helpful discussions with F. Gebhard, L. Tapaszt\'o and P. 
Vancs\'o.
This work was supported in part by the
National Research, Development and Innovation Office (NKFIH) through Grant Nos.~K120569, NN110360 and within the Quantum Technology National
Excellence Program (Project No. 2017-1.2.1-NKP-2017-00001). I.H. is supported by the Alexander von Humboldt Foundation.}

\bibliography{paper_graphene} 

\end{document}